\documentclass[10pt,journal]{IEEEtran}
\usepackage{cite}
\usepackage{amsmath,amssymb,amsfonts}
\usepackage{graphicx}
\usepackage{comment}
\usepackage{multirow}
\usepackage{subcaption}
\usepackage{textcomp}
\usepackage{xcolor}
\usepackage{float}
\usepackage{booktabs}
\usepackage{algorithm}
\usepackage{algpseudocode}
\usepackage{verbatim}
\usepackage{url}
\usepackage{array}
\usepackage{amsthm}
\usepackage{amssymb}
\usepackage{subfiles}
\usepackage{bm}
\algrenewcommand{\algorithmicrequire}{\textbf{Input:}}
\algrenewcommand{\algorithmicensure}{\textbf{Output:}}
\def\BibTeX{{\rm B\kern-.05em{\sc i\kern-.025em b}\kern-.08em
    T\kern-.1667em\lower.7ex\hbox{E}\kern-.125emX}}

\begin{document}

\title{Seamless Handover in Direct-to-Device Satellite Networks: From an Interference-Aware Perspective}


\author{Ye~Li,
		~Yi~Zhou,
		~Yingdong~Hu,
		~Zhe~Ji,
		~Jiaxi~Zhou,
        ~Longyu~Zhou,
		~Yuchen~Liu,
		and~Tony~Q.~S.~Quek~\IEEEmembership{Fellow,~IEEE}

\thanks{Y. Li, Y. Zhou, Y. Hu are with the School of Information Science and Technology, Nantong University, Nantong 226019, China (e-mail: yeli@ntu.edu.cn, zhouyi@stmail.ntu.edu.cn, huyd@ntu.edu.cn).}
\thanks{Z. Ji is with the School of Information and Communication Engineering, Beijing University of Posts and Telecommunications, Beijing 100876, China (e-mail: jiz18@bupt.edu.cn).}
\thanks{J. Zhou is with the Research Center of 6G Mobile Communications, School of Cyber Science and Engineering, Huazhong University of Science and Technology, Wuhan 430074, China (e-mail: zhoujiaxi@hust.edu.cn).}
\thanks {L. Zhou is with Singapore Innovation Research Institute, 068913, and also with Singapore University of Technology and Design, 487372, Singapore (e-mail: zhoulyfuture@outlook.com).}
\thanks{Yuchen Liu is with the department of Computer Science at North Carolina State University, USA (e-mail: yliu322@ncsu.edu).}
\thanks{T. Q. S. Quek is with Singapore University of Technology and Design, Singapore 487372, and also with the Department of Electronic Engineering, Kyung Hee University, Yongin 17104, South Korea (e-mail: tonyquek@sutd.edu.sg).}
\thanks{Corresponding Author: Yingdong Hu.}
}

\maketitle

\begin{abstract}
The Direct-to-device (D2D) satellite network is an important 6G evolution direction to enable seamless ubiquitous connectivity. However, the network faces critical handover challenges due to high satellite mobility and wide beam footprints. Conventional handover strategies, mostly designed for terrestrial networks, may encounter excessive co-channel interference (CCI) and service degradation in the satellite environments. To address the issue, this paper introduces a novel handover optimization method to perform dynamic adjustment of an important parameter called elevation angle threshold (EAT) from an  interference-aware perspective. Explicitly, we first analyze the trade-off between satellite visibility and CCI. Then, we propose a numerical algorithm to determine the optimal EAT that can achieve seamless coverage with CCI. We validate our method using a customized D2D LEO satellite network simulator in the Network Simulator (NS-3). The results demonstrate that the EAT optimization significantly reduces packet loss and hence enhances handover reliability. The work highlights the importance of interference-aware handover design for improving service continuity in future D2D satellite networks.
\end{abstract}


\section{Introduction}
To achieve ubiquitous and seamless connectivity, the terrestrial networks (TNs) have been suffering from coverage gaps in remote areas, oceans, and disaster-stricken scenarios, due to geographical conditions as well as the high cost of infrastructure deployment. To address this challenge, direct-to-device (D2D) satellite networks have emerged as a new solution\cite{LEO2021Z.GAO}. The D2D technology enables hand-held devices to establish direct links with satellites by leveraging satellite-mounted base-station (BS) together with the advances of chips and antennas in mobile devices. The technology fundamentally relaxes traditional satellite communication's reliance on heavy and specialized equipment\cite{D2D}. As a major milestone, the D2D has been documented in 3GPP TR 38.821\cite{TR38.821}, which defines non-terrestrial network (NTN) framework, thereby evolving towards standardized TN and NTN coordination/integration.

However, conventional HO strategies developed for TN rely heavily on pronounced signal strength variations at cell boundaries, and handover is not frequent due to the slow movements of users relative to the BSs. However, this situation drastically changes in D2D scenarios, where satellites move approximately 7.56 km/s relative to the ground. As a result, a user must be handed over to a new satellite or beam within a time scale of every few minutes. Meanwhile, the large beam footprints result in extensive service overlap regions, where signal strength differences among candidate satellites are marginal, making it difficult for user devices to identify the optimal HO target based solely on received signal metrics\cite{TR38.821}. As a consequence, directly applying conventional terrestrial HO mechanisms would be prone to ping-ponging between BSs or HO to suboptimal links, and ultimately cause degraded service continuity and quality. To address these challenges, moving beyond conventional passive, signal-reactive HO schemes toward predictive and decision-oriented mechanisms is essential \ cite {seamlessHO}. In this regard, HO, leveraging the inherent predictability of orbits, along with user geographical information and network load conditions, is key, leading to the regime of conditional HO (CHO).

The seamlessness of HO largely depends on the number of devices' visible satellites. Such a number is determined by an important term called the elevation angle threshold (EAT) of the satellites, which is the minimum angle above the horizon at which communication between the device and the satellite is considered viable. The lower EAT corresponds to more visible satellites. In current practice with sparse satellite users, it is natural to set the EAT in the user device on the basis that as many satellites as possible are visible. In D2D scenarios with much denser user distribution, however, a more visible satellite may also mean more co-channel interference (CCI). The CCI can be from a beacon or a paging signal, and can also be data signals if spectrum reuse exists between neighboring satellites. When a user is located within overlapping coverage regions of multiple satellites, CCI from non-serving satellites can significantly degrade the signal-to-interference-plus-noise ratio (SINR) of the serving link \cite{LEOInterferenceGAO,inter-satCCI}, and hence affect HO measurement and signaling to threaten service continuity. \textbf{Such a trade-off between the number of visible satellites and interference has not been sufficiently investigated in the existing literature.} To address this issue, this paper studies interference-aware EAT optimization in the D2D scenario. The main contributions are summarized as follows.

\begin{itemize}
    \item We propose to formulate the handover problem in D2D satellite networks by explicitly balancing satellite visibility and co-channel interference (CCI) through elevation angle threshold (EAT) optimization.
    \item We propose a numerical algorithm to dynamically determine the optimal EAT, ensuring seamless coverage while minimizing CCI based on constellation configuration and user location.
    \item We develop a customized simulator in ns-3 for LEO D2D networks and demonstrate that an optimized EAT can significantly reduce packet loss and hence enhance handover seamlessness.
\end{itemize}

The remainder of this paper is organized as follows: Section II reviews the existing HO strategies; Section III analyzes the trade-off between satellite coverage and CCI, and formulates the optimal EAT problem; In section IV, a D2D LEO satellite network simulation platform is constructed; In section V, we conduct experiments to validate the proposed scheme; Finally, section VI concludes the paper and outlines future research directions.

\section{Overview of Satellite Handover Strategies}
This section provides a systematic overview of these strategies based on their mechanisms, and discusses their performance implications in D2D scenarios.

\subsection{Measurement-based Handover}
The analysis in 3GPP TR 38.821 indicates that the application of measurement-based HO strategies from TN to LEO satellite networks faces limitations. These strategies rely on instantaneous measurements, such as RSRP or SINR, and assume the existence of pronounced signal strength gradients near cell boundaries. This precondition, however, may change in LEO satellite networks characterized by wide beam footprints. When a user traverses from the center to the edge of a satellite beam, the variation in received signal strength may be as small as approximately 3 dB, which is of an order of magnitude much lower than that with TN BSs, which is commonly up to 50 dB\cite{seamlessHO}. Such faint signal variations, if further combined with potential CCI experienced by user equipment (UE) in beam overlap regions, would complicate the task of distinguishing between serving and neighboring cells based solely on signal strength.

Building upon this observation, Juan et al.\cite{juan2022handover} conducted a comparative study using 5G NR HO as a baseline and evaluated several enhanced schemes, including location-based CHO and antenna-gain-based CHO, with respect to HO reliability, resource utilization, signal quality, and signaling overhead. Their results show that directly reusing measurement report triggering strategies based on events such as A3 may increase the risk of HO ping-ponging or suboptimal HO decisions, which would degrade HO robustness and service continuity. These findings collectively suggest that, in LEO NTN scenarios, the integration of multi-weight CHO and ephemeris-based predictive HO may offer a viable means to augment, or partially replace, traditional measurement-driven decision frameworks under certain conditions.

\subsection{Conventional Conditional Handover}

Originally introduced by 3GPP as an enhancement for TN, CHO decouples the preparation and execution phases of HO. Specifically, based on early measurement reports, the network preconfigures multiple candidate cells and one or more execution conditions for the UE. The UE thus holds the HO command in advance and autonomously triggers access to the target cell only when the predefined execution conditions are satisfied. This mechanism is designed to mitigate radio link failures (RLFs) caused by the inability to deliver HO commands in time under fast channel degradation.

Early evaluations of directly applying CHO to NTN scenarios have confirmed its effectiveness in reducing RLFs, stemming from late HO execution. However, subsequent studies also revealed its limitations. If the triggering and/or execution conditions of the CHO schemes rely on instantaneous measurements, the fundamental challenges of flat signal gradients and interference-prone measurements in NTN environments remain an issue. Research indicates that while CHO can suppress RLFs, it may simultaneously introduce a substantial increase in unnecessary HOs and ping-ponging, with HO ratios exceeding 60 percent in some scenarios\cite{juan2022performance}. Essentially, the conventional CHO schemes improve the reliability of HO execution, but not enhance the quality of the information driving the HO decisions. When the information is noisy, ambiguous, or delayed, CHO schemes may even trigger excessive signaling exchanges, increasing the risk of the so-called ``signaling storms'' \cite{zhang2024signalling}.

\subsection{Predictive and Ephemeris-based Handover}

Given that solely refining the execution mechanism with CHO may not be sufficient in NTN, a line of recent studies has explored the incorporation of predictive information into the CHO framework. In particular, predictive HO schemes based on satellite ephemeris and user location have been proposed to improve decision accuracy by leveraging the deterministic and periodic nature of satellite motion \cite{seamlessHO}. 3GPP TS 38.331 has formally introduced location-based (Event D1) and time-based (Event T1) triggering conditions\cite{TS38.331}, which are designed to operate in conjunction with measurement-based strategies. Existing studies further confirm that incorporating ephemeris-based prediction logic into the CHO framework can reduce RLFs and overall HO ratios, outperforming measurement-based HO strategies\cite{lee2023location}.

For instance, \cite{li2023ephemeris} proposed a live migration scheme cooperating ephemeris and UE handover requests, which exploits satellite orbital predictability from ephemeris to pre-transfer user and network function contexts, ensuring seamless service continuity with migration latency controlled within seconds. In \cite{saglam2023conditional}, a 3GPP-compatible conditional handover (CHO) algorithm for NTNs was developed, where ephemeris data is integrated with location/time-based triggers, and multi-tier candidate cells and group HO techniques were adopted to mitigate signaling overhead.

\section{CCI-Aware CHO via EAT Optimization}
Despite the diverse CHO implementations for NTN, one limitation with the existing CHO schemes is that they can not take the potential co-channel interference (CCI) into account. In terrestrial cellular systems, when adjacent cells reuse frequency resources, UE located near cell edges is exposed to CCI originating from either the data transmissions of neighboring BSs or the continuously broadcasting control and reference signals. In LEO satellite constellations, the characteristics of CCI is fundamentally reshaped by the wide beam footprints of satellites and the dense constellation deployment, which give rise to extensive coverage overlap regions on the ground. UEs located within these regions inevitably receive signals from multiple visible satellites while being served by a target satellite. As a result, CCI in LEO D2D networks can be jointly contributed by all non-serving satellites. These satellites are geometrically visible to the UE, either being their control signals or downlink data transmissions to other users sharing the same spectrum. Ignoring such CCI during CHO can significantly mislead the HO decisions.

\begin{figure}[t]
\centering
\includegraphics[width=1\columnwidth]{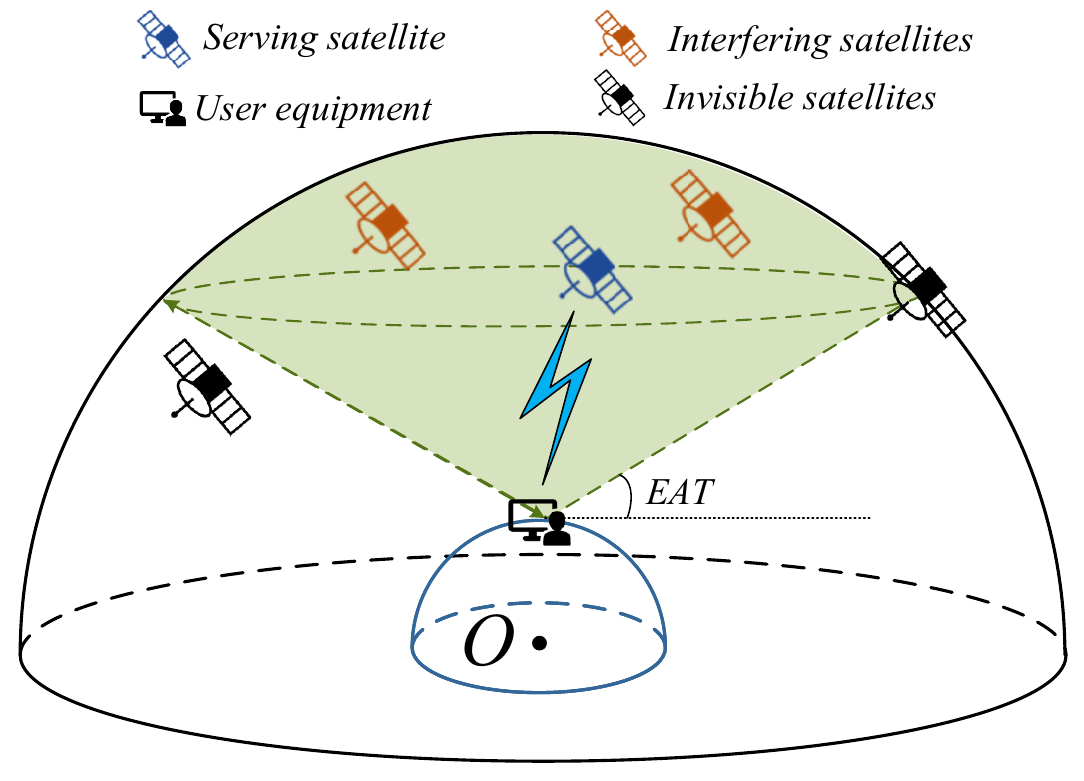}
\caption{Illustration of the EAT.}
\label{fig1}
\end{figure}

Reducing CCI requires controlling the number of satellites these are simultaneously visible from the perspective of each UE. As illustrated by the green region in Fig.~\ref{fig1}, it can be achieved by dynamically adjusting the EAT. The EAT represents the minimum elevation angle at which a UE can establish and maintain a reliable communication link with the satellite. Given EATs of the satellites, the extent of footprint overlap between them, and hence the CCI, is determined. On the other hand, the coverage continuity of the satellite network is also determined. Therefore, the EAT serves as a pivotal control point in balancing two conflicting objectives. From a coverage perspective, a lower EAT admits a larger set of visible satellites and increases the number of candidate links, thereby reducing the probability of service interruption. From an interference perspective, however, this expanded visibility simultaneously introduces more potential CCI sources, which may degrade the received SINR and impair link reliability.

Existing studies and constellations in practice typically consider EAT configuration based on the requirements of continuous coverage of satellites, without explicitly considering potential CCI that may be caused by more visible satellites. In a constellation, the number of visible satellites is unevenly distributed across different latitude regions, and a denser distribution is often observed in mid- to high-latitude regions. Given this fact, using the same low EAT to ensure seamless coverage in low-altitude area would cause higher CCI in high-latitude area, and using a high EAT would otherwise comprise the coverage in-latitude area. While current constellations (e.g. Starlink) may select different fixed EATs based on latitude ranges, no explicit interaction between the CCI and latitude has been taken into account. Furthermore, note that even for users on the same latitude, whether they are located right beneath the satellite orbit or between adjacent orbits would also affect the number of visible satellites given the same EAT, meaning that the CCI level could be different. As a result, the current approach of configuring EAT solely based on latitude is insufficient to optimally balance CCI and coverage.

These observations underline the importance of dynamically selecting an optimal EAT based on known constellation parameters as well as the location of the UE, with the objective of identifying a feasible EAT region that balances coverage and CCI. To this end, we propose a numerical approach aimed at identifying the Pareto-optimal boundary between coverage continuity and CCI suppression. Explicitly, with satellite ephemeris, user location, a set of candidate elevation thresholds, our algorithm evaluates each candidate EAT by computing the set of visible satellites over time. The CCI level is then quantified according to the interference models corresponding to the specific scenarios (e.g. antenna and beam hopping pattern, spectrum reuse strategy, etc). The framework searches for the optimal EAT that maximizes a system‑level objective function, which inherently balances service continuity (ensuring at least one satellite is visible) against the incurred interference. As concrete examples, in the special zero‑interference scenario the algorithm would reduce to maximizing the elevation threshold under a coverage constraint for less propagation loss, whereas in a complete frequency‑reuse scenario the algorithm would lead to finding EATs corresponding to exactly one visible satellite.

\section{A Simulator for CCI-Aware CHO Evaluation}
To evaluate HO performance under CCI constraints, in this section, we construct a satellite network simulator based on ns-3. The simulator captures essential components of a LEO D2D scenario, including constellation configuration, satellite mobility, NTN propagation loss model, and BS-on-satellite with 3GPP-compliance protocol stack.

\begin{figure}
\centering
\includegraphics[width=1\columnwidth]{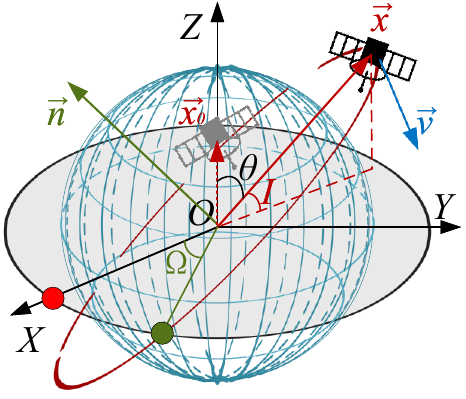}
\caption{Satellite constellation architecture under ITRF.}
\label{fig2}
\end{figure}

\subsection{Satellite Constellation and Mobility Model}
To construct a simulation platform that closely reflects realistic operating conditions, we first develop a geometric and kinematic model of the LEO satellite constellation, where the International Terrestrial Reference Frame (ITRF) is adopted as a unified spatiotemporal reference, as shown in Fig.\ref{fig2}. In this model, the Earth is simplified as a sphere with a radius of 6371km, and the Earth center is taken as the origin of the coordinate system. The X-axis lies in the equatorial plane and points toward the vernal equinox, the Z-axis coincides with the Earth's rotation axis and points toward the North Pole, and the Y-axis is orthogonal to the XOZ plane, together forming a right-handed Cartesian coordinate system. We assume that all satellites move along circular orbits. The spatial orientation of each orbital plane is characterized by two key parameters: the orbital inclination, which defines the angle between the orbital plane and the Earth's equatorial plane, and the right ascension of the ascending node (RAAN), which specifies the initial orientation of the orbital plane in space. Furthermore, we assume that satellites are uniformly distributed within each orbital plane. The relative phasing between satellites in adjacent orbital planes is controlled by a phase factor. For Walker constellations, as exemplified by Starlink, a phase factor is introduced to regulate this inter-plane offset. When the factor is zero, satellites across all orbital planes are phase-aligned, whereas a factor of one yields a more uniform global coverage distribution. By adjusting these key constellation parameters, the proposed simulator is capable of flexibly reproducing a wide range of mainstream as well as customized LEO constellation architecture.

We use the following mobility model to determine satellite position over the time. As shown in Fig.\ref{fig2}, the satellite's orbital motion can be described using the orbital velocity, the instantaneous position vector, the initial position vector , and the unit normal vector of the orbital plane. Furthermore, the satellite mean anomaly can be expressed as the angle swept by the satellite orbiting at a constant angular speed. Therefore, the satellite instantaneous position vector can always be obtained using Rodrigues' formula based on the above parameters.

\subsection{A System-level Simulation}
\begin{figure}[t]
\centering
\includegraphics[width=1\columnwidth]{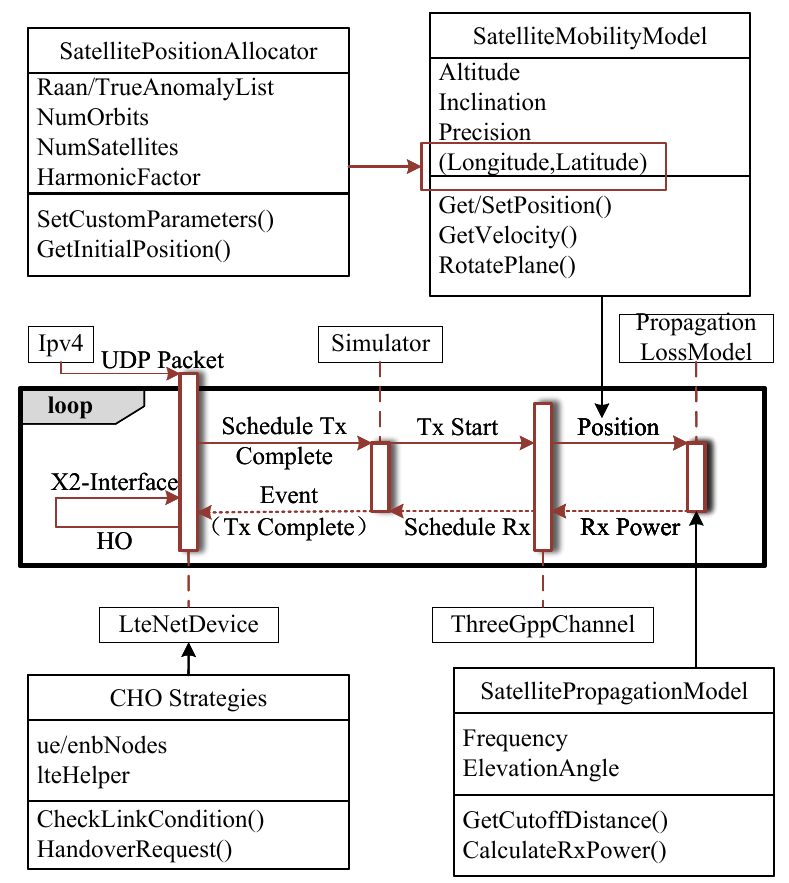}
\caption{D2D LEO satellite network simulator.}
\label{fig3}
\end{figure}

The above constellation and mobility models are implemented in NS-3 illustrated in Fig. \ref{fig3}. The implementation inherits and extends ns-3's base mobility model. To simulate BS on satellites, we utilize ns-3's built-in \texttt{lte} module, which is long-time verified. The module features a 3GPP-compliance LTE protocol stack, and has built-in interference models. To adapt to the NTN scenario, the propagation loss model which complies with 3GPP TR38.811 is adopted to replace the original terrestrial loss model in \texttt{lte}\cite{sandri2023implementation}. 

More specifically, we have implemented the following customized modules corresponding to the models described in section IV.A: \texttt{SatellitePositionAllocator} configures orbital parameters and initial phases of satellites, \texttt{SatelliteMobilityModel} simulates continuous orbital movement, and \texttt{ThreeGppNTNPropagationLossModel} specifically emulates large-scale fading characteristics of satellite-to-ground links, calculating received power based on parameters such as frequency, shadowing and building penetration losses, thereby providing an accurate propagation environment for link-level simulation.

To assess CHO, a specific interface has been added to \texttt{LteNetDevice}, which can directly control \texttt{lte}'s X2 interface. This allows to preconfigure target satellites so that the UE can autonomously initiate handover once specific conditions are satisfied. As an example, we have implemented the closest satellite handover (CSHO)\cite{chowdhury2006handover} strategy in this work. Under CSHO, the UE always connects to a visible satellite that is geometrically closest to its current location. To support the scheme, dedicated monitoring functions are added to continuously track the real-time satellite-to-ground distances and satellite visibility. Once the predefined handover condition is met, a handover is triggered via \texttt{LteHelper::HandoverRequest}. The handover request further invokes \texttt{DoHandoverRequest}, where the RRC layer of the source satellite initiates handover signaling to the target satellite through the \texttt{SendHandoverRequest} method over the X2-interface, thereby enabling seamless satellite migration.

\begin{figure}[t]
\centering
\includegraphics[width=1\columnwidth]{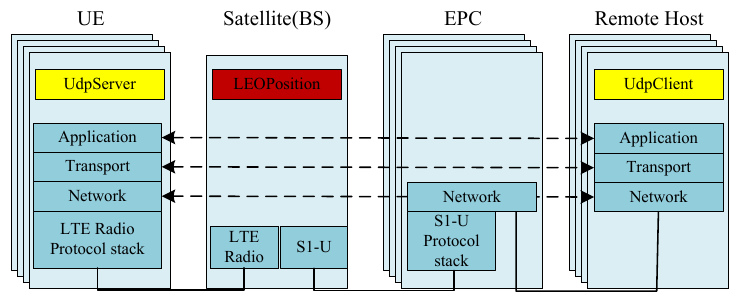}
\caption{End-to-end evaluation of D2D satellite network.}
\label{fig4}
\end{figure}

Given the above framework, UDP-based packet transmitters and receivers can then be deployed to evaluate the CHO performance from an end-to-end transmission perspective as shown in Fig. \ref{fig4}, which establishes a data path that traverses the evolved packet core (EPC) and BSs (as satellites), thereby emulating the complete downlink data flow over a dynamic constellation.

\section{Simulation Results And Discussion}

In this section, we demonstrate how CCI-aware EAT choice may affect CHO performance by evaluating the EAT determined by the proposed Algorithm 1 in the constructed system-level simulator. We perform the simulation based on the publicly available parameters of Starlink Phase I constellation. The constellation follows a Walker architecture, consisting of 72 orbital planes with 22 satellites deployed per plane. The orbital inclination is set to 53 degree, the satellites operate at an altitude of approximately 550 km. If not otherwise specified, the default EAT setting is 25 degree. We set the TxPower, Antenna Gain, Bandwidth, UDP Packet Rate, and Frequency as 20w, 70.2dBi, 4.5MHz, 8.192Mbps, 11.7GHz, respectively. The results can be extended to other interference models as well.

We first validate the core insight established in Section III that the optimal EAT depends on the latitude. To achieve this, we fix the user longitude at 0 degree and vary the latitude from 0 to 60 degrees, with a sampling resolution of 0.5. For each latitude, Algorithm 1 is executed to determine the optimal EAT. It is not difficult to see that, under the adopted interference model, the optimal EAT is just the maximum EAT that ensures exactly one visible satellite (to ensure seamless coverage).

\begin{figure}[t]
\centering
\includegraphics[width=\columnwidth]{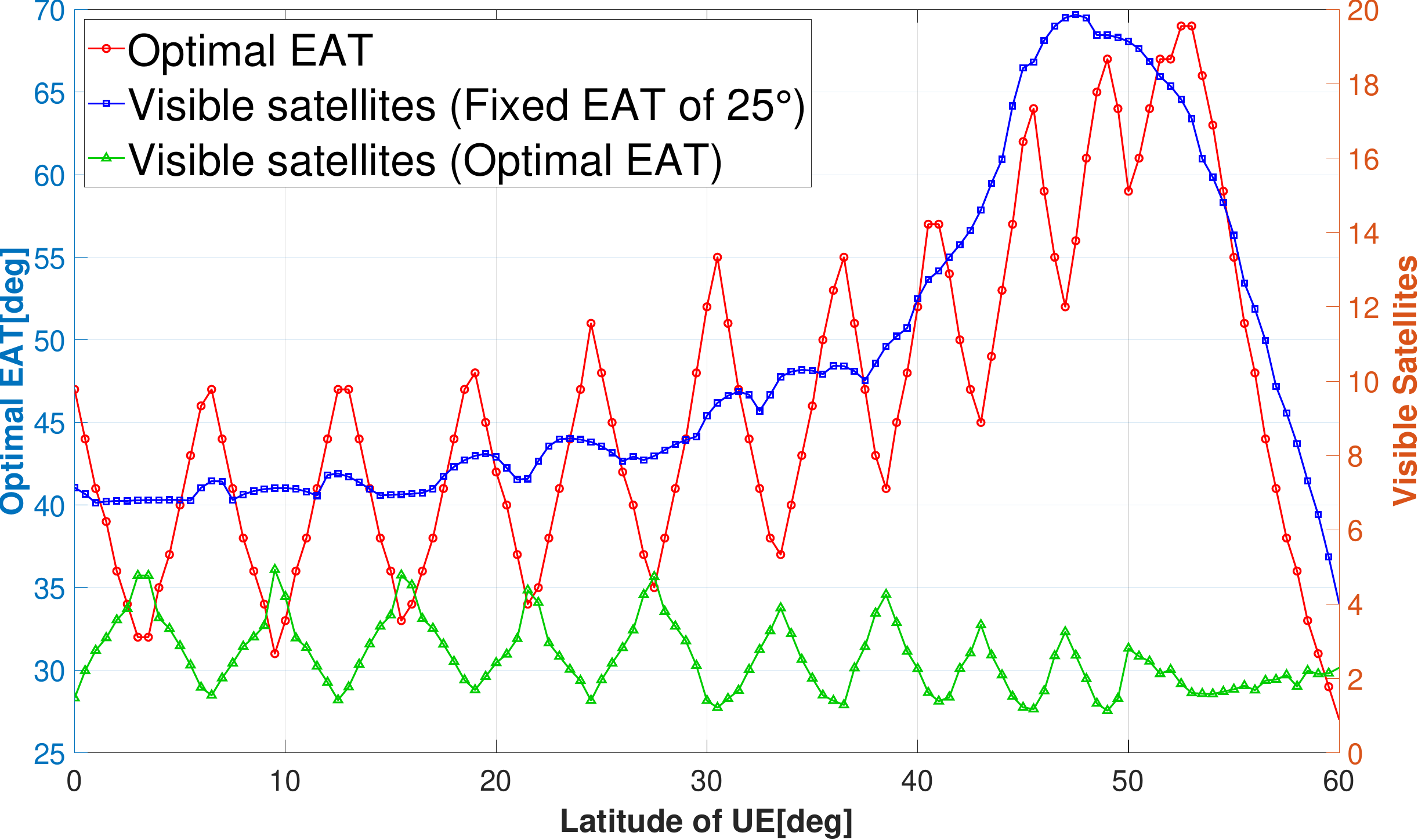}
\caption{EAT and visible satellites for seamless coverage at different latitudes.}
\label{fig5}
\end{figure}

As shown in Fig. \ref{fig5}, the numerically determined optimal EAT exhibits an overall increasing trend with latitude. The figure also shows the numbers of visible satellites of using fixed and the optimal EATs, respectively. At higher latitudes, satellite orbits become denser in the sky, requiring a higher EAT to effectively filter out more low-elevation satellites and thereby avoid prolonged exposure to strong CCI from them. This helps stabilize handover measurements and reduces the likelihood of ambiguous or unnecessary handover decisions. On the other hand, the curve also shows a periodic oscillation pattern. We note that the peaks of these oscillations correspond to latitudes located directly beneath the ground projection of satellite orbital planes. In these regions, the coverage redundancy is high, demanding for the adoption of a larger EAT to avoid CCI. In contrast, the downs occur at latitudes situated in the intermediate regions between adjacent orbital plane projections. There, the coverage redundancy is reduced, and a lower EAT becomes necessary to admit more satellites to ensure coverage. At the highest latitudes, a sharp decline in the optimal EAT is observed, which is attributed to fewer and fewer visible satellites due to the orbital inclination limit (i.e., 53 degree in this constellation). As the user latitude exceeds a limit, the number of visible satellites decreases rapidly, where handover performance becomes increasingly coverage-constrained rather than interference-limited.

\begin{figure}[t]
\centering
\includegraphics[width=\columnwidth]{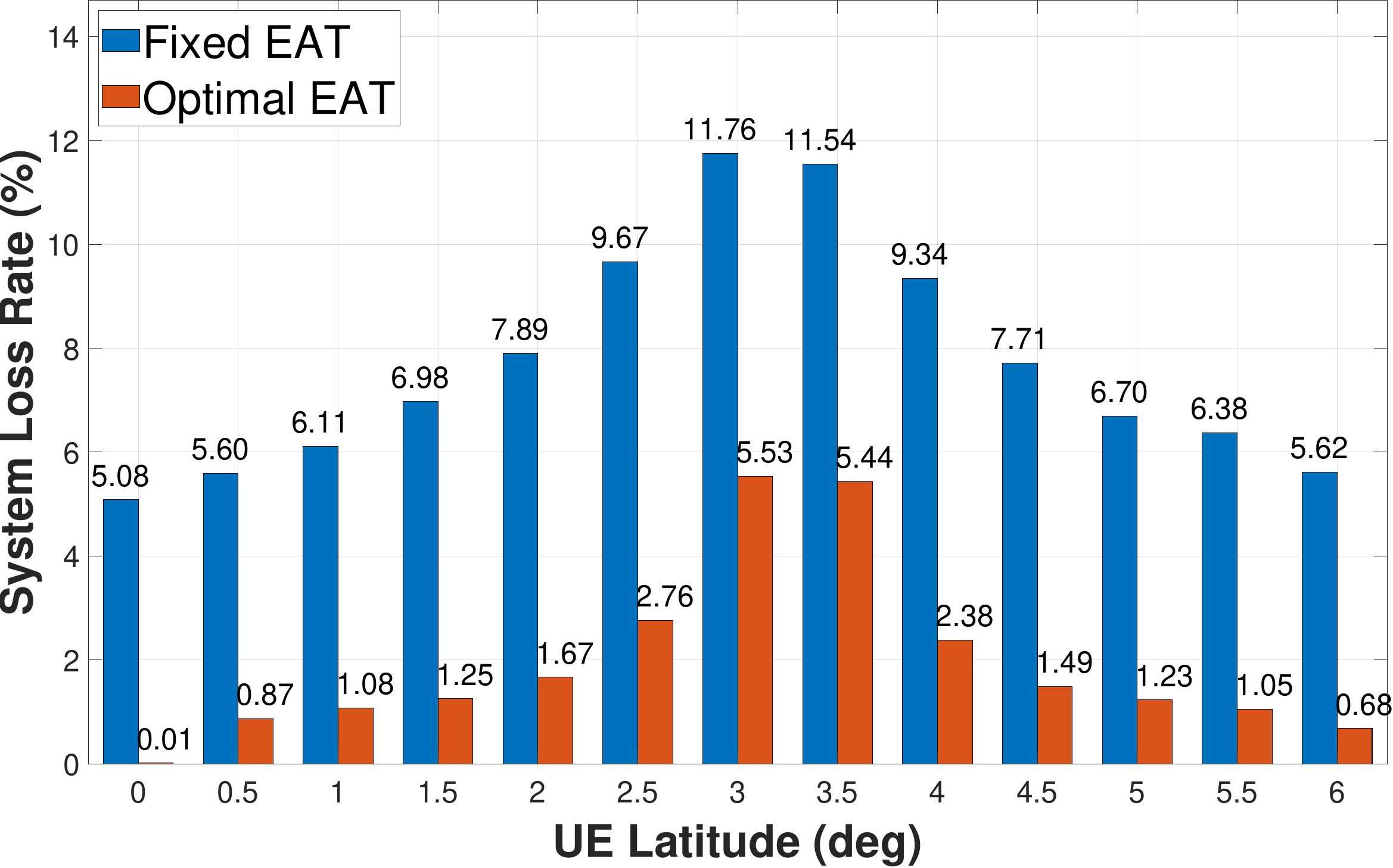}
\caption{Comparison of loss rate using fixed and optimal EATs.}
\label{fig6}
\end{figure}

To further demonstrate the importance of EAT optimization, we conduct a comparative study using the proposed simulator within a smaller region located between the ground projection of two adjacent orbits. In the experiment, the user longitude is fixed, while its latitude is narrowly varied from 0 to 6 degrees in increments of 0.5. In all the simulations, CSHO strategy is adopted for HO. Each simulation runs UDP transmissions to the UE for 200s, during which the UE typically experiences one to three satellite handovers. For a given latitude, both the fixed-EAT baseline and the proposed optimal-EAT scheme undergo the same number of handovers, ensuring that the observed performance differences are attributed to interference mitigation rather than handover frequency. To measure the HO performance, we record the system end-to-end packet loss rate under two configurations, which directly reflects the service continuity. The results, illustrated in Fig. \ref{fig6}, show that the optimal EAT strategy (orange bars) consistently achieves a significantly lower packet loss rate than the fixed-EAT baseline. These results imply that wisely setting EAT can effectively suppress CCI while preserving service continuity to achieve seamless HO, enhancing the end-to-end data delivery performance in LEO D2D networks.

\begin{figure}[t]
\centering
\includegraphics[width=\columnwidth]{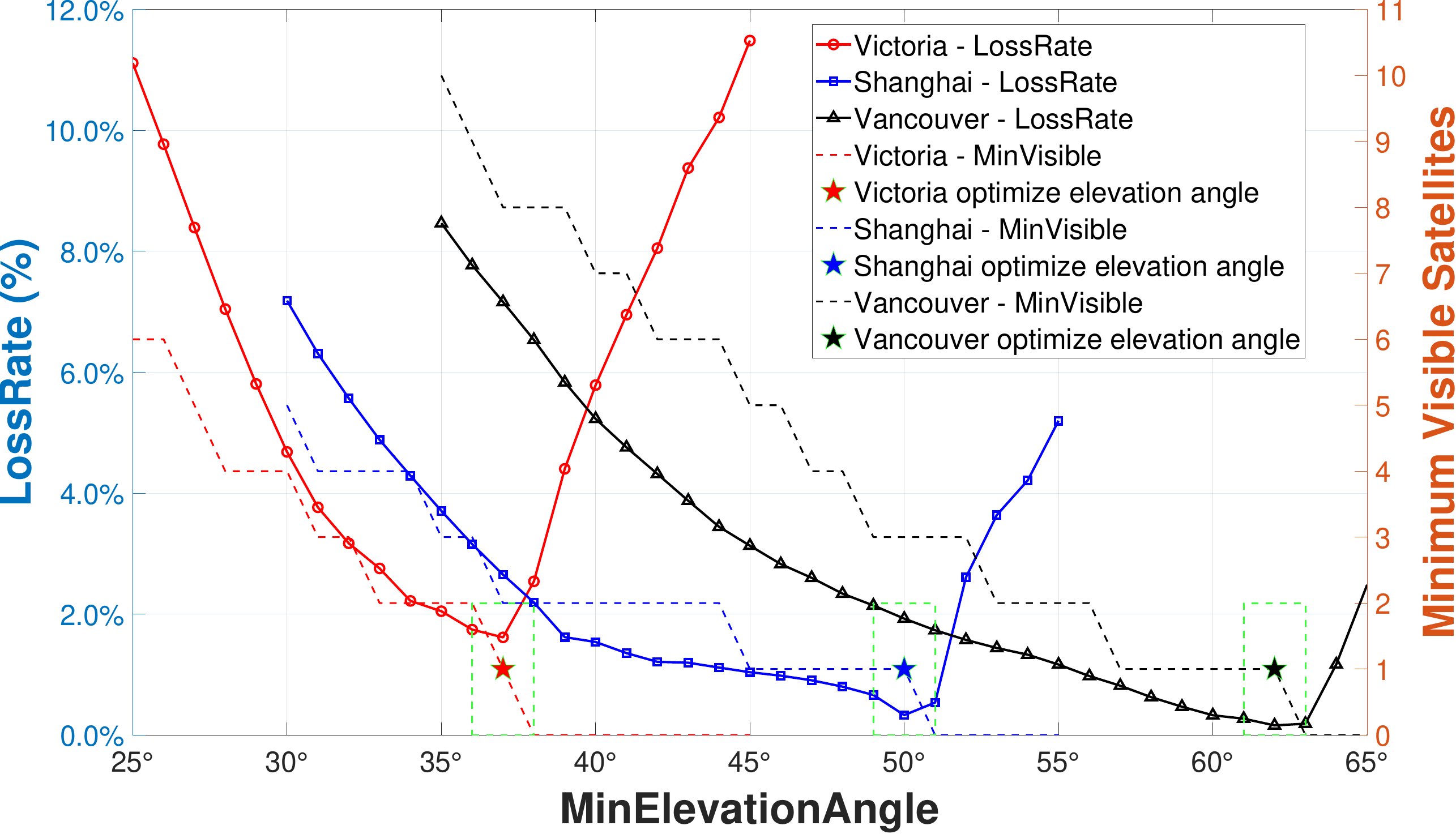}
\caption{Comparison of numerical evaluation and ns-3 simulation results.}
\label{fig7}
\end{figure}

Finally, we conduct long-term traffic simulations at three representative low-, mid-, and high-latitudes, namely Victoria, Seychelles (-4.3N, 55.27E), Shanghai, China (31.23N, 121.47E), and Vancouver, Canada (49.13N, -123.06E), respectively, to assess the end-to-end performance. For each UE location, we simulate using various EATs that are progressively increased, and the end-to-end packet loss rate over the entire communication session is monitored. Across all the three latitudes, the packet loss curves exhibit a clear U-shaped behavior as a function of EAT. When the EAT is set low, service continuity is preserved, but severe CCI results in high error rates and substantial packet loss. As the EAT increases toward the optimum point predicted by our algorithm (marked as stars), the packet loss rate reaches its minimum, indicating an optimal balance between coverage and interference suppression. Further, when the EAT increases beyond the optimal value, the probability of service interruption increases, and packet losses caused by coverage outages begin to dominate, leading to a degradation in overall performance.

\section{Conclusion}
In this study, we constructively discuss, analyze, and optimize the handover process in D2D Satellite networks from the perspective of interference management. We identify the fundamental role of EAT in balancing between CCI and seamless coverage. Based on a numerical algorithm for optimal EAT under certain interference models and a customized system-level LEO D2D satellite network simulator, we demonstrate the end-to-end packet loss rate is reduced using the optimized EATs. This insight offers a new system design perspective for improving service quality in the D2D satellite networks. The simulation platform also provides a useful means for designing/evaluating other CHO mechanisms under various interference models. In the future, extending the work to integrating with a full 3GPP NTN compatible protocol stack is also an interesting work to pursue.


\begin{thebibliography}{10}
	
	\bibitem{LEO2021Z.GAO}
	Shicong Liu, Zhen Gao, Yongpeng Wu, Derrick Wing~Kwan Ng, Xiqi Gao, Kai-Kit
	Wong, Symeon Chatzinotas, and Bj{\"o}rn Ottersten.
	\newblock {LEO} satellite constellations for {5G} and beyond: {How} will they
	reshape vertical domains?
	\newblock {\em IEEE Communications Magazine}, 59(7):30--36, 2021.
	
	\bibitem{D2D}
	Chang Pei, Wang Wei, Zhao Yijun, Wang Yingkui, Lin Xinfa, Gan Wei, et~al.
	\newblock Analysis of the current situation, evolution, and policy of
	direct-to-handset satellite.
	\newblock {\em The Frontiers of Society, Science and Technology}, 7(4), 2025.
	
	\bibitem{TR38.821}
	TR~38.821.
	\newblock Solutions for {NR} to support non-terrestrial networks {(NTN)}.
	\newblock {\em 3rd Generation Partnership Project (3GPP), v16.2.0}, Mar.2023.
	
	\bibitem{LEOInterferenceGAO}
	Chong Xu, Feng Liu, Junyi Yang, Yafeng Ma, Zhen Gao, Zhenyu Xiao, and Xiang-Gen
	Xia.
	\newblock {MIMO}-based multi-{LEO}-satellite cooperative grant-free random
	access for {IoT} massive connectivity.
	\newblock {\em IEEE Transactions on Wireless Communications},
	24(12):10644--10659, 2025.
	
	\bibitem{inter-satCCI}
	Akram Al-Hourani.
	\newblock An analytic approach for modeling the coverage performance of dense
	satellite networks.
	\newblock {\em IEEE Wireless Communications Letters}, 10(4):897--901, 2021.
	
	\bibitem{seamlessHO}
	Feng Wang, Dingde Jiang, Zhihao Wang, Jianguang Chen, and Tony~QS Quek.
	\newblock Seamless handover in {LEO} based non-terrestrial networks: {Service}
	continuity and optimization.
	\newblock {\em IEEE Transactions on Communications}, 71(2):1008--1023, 2022.
	
	\bibitem{juan2022handover}
	Enric Juan, Mads Lauridsen, Jeroen Wigard, and Preben Mogensen.
	\newblock Handover solutions for {5G} low-earth orbit satellite networks.
	\newblock {\em IEEE Access}, 10:93309--93325, 2022.
	
	\bibitem{juan2022performance}
	Enric Juan, Mads Lauridsen, Jeroen Wigard, and Preben Mogensen.
	\newblock Performance evaluation of the {5G} {NR} conditional handover in
	{LEO-based} non-terrestrial networks.
	\newblock In {\em 2022 IEEE Wireless Communications and Networking Conference
		(WCNC)}, pages 2488--2493. IEEE, 2022.
	
	\bibitem{zhang2024signalling}
	Bohan Zhang, Mohammad~A Salahuddin, Peng Hu, Yunli Wang, Noura Limam, Bo~Sun,
	Diogo Barradas, and Raouf Boutaba.
	\newblock Signalling load-aware conditional handover in {5G} non-terrestrial
	networks.
	\newblock In {\em 2024 20th International Conference on Network and Service
		Management (CNSM)}, pages 1--9. IEEE, 2024.
	
	\bibitem{TS38.331}
	TS~38.331.
	\newblock Radio resource control {(RRC)} protocol specification.
	\newblock {\em 3rd Generation Partnership Project (3GPP), v18.7.0}, Sep.2025.
	
	\bibitem{lee2023location}
	Jongtae Lee, Wonjae Lee, and Jae-Hyun Kim.
	\newblock Performance evaluation of location-based conditional handover scheme
	using {LEO} satellites.
	\newblock In {\em 2023 14th International Conference on Information and
		Communication Technology Convergence (ICTC)}, pages 1642--1644. IEEE, 2023.
	
	\bibitem{li2023ephemeris}
	Jinze Li, Xi~Chen, Shengfeng Wang, Changsheng Zhou, Chengli Mei, Ping Du, Hao
	Qin, and Xinsheng Zhao.
	\newblock An ephemeris and handover based live migration scheme for seamless
	service transfer in {LEO} communications system.
	\newblock In {\em 2023 3rd International Conference on Intelligent
		Communications and Computing (ICC)}, pages 98--102. IEEE, 2023.
	
	\bibitem{saglam2023conditional}
	Mehmet~Izzet Saglam.
	\newblock Conditional handover for non-terrestrial networks.
	\newblock In {\em 2023 10th International Conference on Wireless Networks and
		Mobile Communications (WINCOM)}, pages 1--5. IEEE, 2023.
	
	\bibitem{sandri2023implementation}
	Mattia Sandri, Matteo Pagin, Marco Giordani, and Michele Zorzi.
	\newblock Implementation of a channel model for non-terrestrial networks in
	ns-3.
	\newblock In {\em Proceedings of the 2023 Workshop on ns-3}, pages 28--34,
	2023.
	
	\bibitem{chowdhury2006handover}
	Pulak~K Chowdhury, Mohammed Atiquzzaman, William~D Ivancic, et~al.
	\newblock Handover schemes in satellite networks: {State}-of-the-art and future
	research directions.
	\newblock {\em IEEE Commun. Surv. Tutorials}, 8(1-4):2--14, 2006.
	
\end{thebibliography}
\end{document}